\documentclass{article}
\usepackage{spconf,amsmath,graphicx,bm,amssymb,booktabs,cite,url}
\usepackage{color}

\title{A Hybrid Discriminative and Generative System for Universal Speech Enhancement}

\name{Yinghao Liu\textsuperscript{1}, Chengwei Liu\textsuperscript{1}, Xiaotao Liang\textsuperscript{1}, Haoyin Yan\textsuperscript{1,2}, Shaofei Xue\textsuperscript{1,2}\sthanks{Corresponding Author.}, Zheng Xue\textsuperscript{1}}
\address{\textsuperscript{1}Intelligent Connectivity, Alibaba Group \\
\textsuperscript{2}Tongyi AI Lab, Alibaba Group \\
\{liuyinghao.lyh, liuchengwei.lcw, xiaotao.lxt, yanhaoyin.yhy, shaofei.xsf, jeremy.xz\}@alibaba-inc.com}

\begin{document}
%
\maketitle
\begin{abstract}
Universal speech enhancement aims at handling inputs with various speech distortions and recording conditions. In this work, we propose a novel hybrid architecture that synergizes the signal fidelity of discriminative modeling with the reconstruction capabilities of generative modeling. 
Our system utilizes the discriminative TF-GridNet model with the Sampling-Frequency-Independent strategy to handle variable sampling rates universally. In parallel, an autoregressive model combined with spectral mapping modeling generates detail-rich speech while effectively suppressing generative artifacts.
Finally, A fusion network learns adaptive weights of the two outputs under the 
optimization of signal-level losses and the comprehensive Speech Quality Assessment (SQA) loss. 
Our proposed system is evaluated in the ICASSP 2026 URGENT Challenge (Track 1) and ranks 
the third place.
\end{abstract}
\section{Introduction}
\label{sec:intro}

Universal speech enhancement (USE) aims to 
recover degraded speech with diverse distortions and recording conditions (e.g., varying sampling rates)~\cite{urgent_overview}.
Recent advances have categorized solutions into discriminative and generative approaches. 
Discriminative models, such as TF-GridNet \cite{tfgridnet}, 
excel at signal fidelity and noise suppression but often struggle to reconstruct severely corrupted speech components~\cite{continuous}.
Conversely, generative models can reconstruct high-quality speech but frequently suffer from hallucinations and artifacts due to imperfect alignment between the learned generative prior and the true underlying clean speech distribution.
To address these issues, we propose a hybrid network for USE, which integrates the strengths of both paradigms.
Specifically, we utilize the TF-GridNet to produce high-fidelity, noise-suppressed estimates and 
employ an autoregressive (AR) module with a spectral mapping \cite{csm} head to 
generate detail-rich but low-hallucination speech.
Finally, a fusion network adaptively combines their outputs, yielding enhanced speech with fine-grained details and reduced artifacts.
The proposed system is evaluated on the Track 1 (USE) of URGENT Challenge, 
obtaining promising performance in terms of both intrusive and non-intrusive metrics.

\section{Method}
\label{sec:method}

\subsection{Discriminative Branch}
We adopt TF-GridNet as the discriminative backbone to process complex spectrograms via a grid of time-frequency blocks. 
To preserve spectral integrity under varying input sampling rates, we apply the Sampling-Frequency-Independent (SFI) \cite{uses} strategy: the Short-Time Fourier Transform (STFT) window and hop durations are fixed, while the number of frequency bins is adjusted according to the sampling rate.

\subsection{Semantic-conditioned Refinement Branch}
\begin{figure}[htb]
    \centering
    \includegraphics[width=\linewidth]{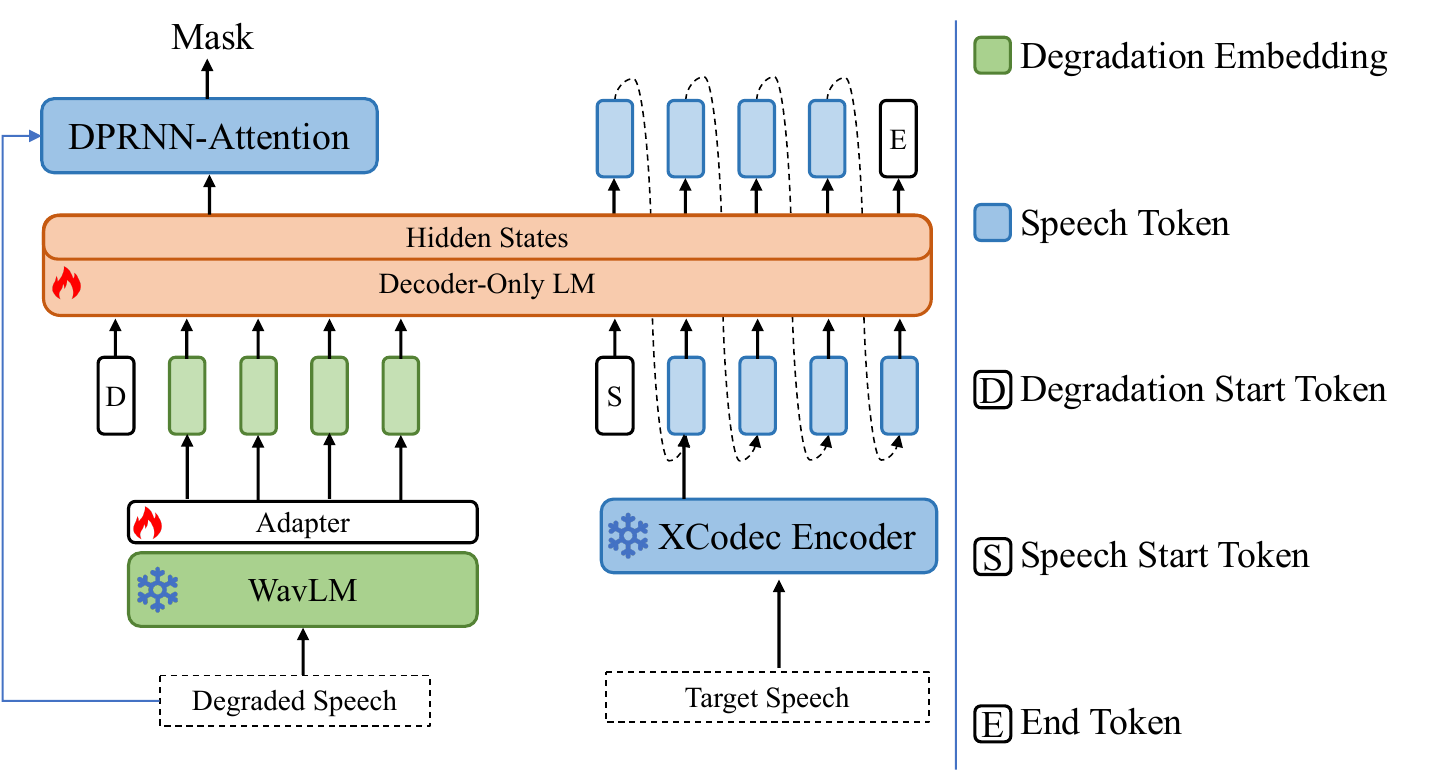}
    \caption{Architecture of the proposed generative branch model.}
    \label{fig:ar_model}
    \vspace{-1em}
\end{figure}

Inspired by \cite{continuous}, 
we utilize the powerful generation capability of AR modeling to deal with complex distortions in USE, as illustrated in Fig.~\ref{fig:ar_model}.
Specifically, we extract robust semantic and acoustic representation from degraded speech using a pre-trained WavLM \cite{wavlm} with a trainable linear adapter. 
This representation serves as conditional prefix, 
leading the decoder-only language model (LM) to autoregressively 
predict discrete tokens of the clean speech, 
which are extracted by the pre-trained X-Codec \cite{xcodec}. 
We only utilize the tokens of the first RVQ \cite{soundstream} layer, because 
they capture the most salient semantic and perceptual information of speech, while higher-layer codes mainly refine low-level details. 

To circumvent the information bottleneck of discrete tokens, we adopt a DPRNN \cite{dprnn} model to directly predict the 
clean speech spectrum by fusing the semantic-rich representations from the last LM layer with the acoustic-rich features derived from the degraded spectrum. 
Specifically, we concatenate the magnitude, real, and imaginary components of the degraded STFT spectrum as the DPRNN input, and apply several convolutional blocks to downsample along the frequency dimension. 
Within each dual-path block, the cross attention is applied after inter-frame LSTM \cite{lstm}, where the representations from LM serve as key matrix and value matrix, 
effectively integrating global semantic guidance with local acoustic nuances. 
Finally, a convolutional decoder up-samples the fused features to the original time-frequency resolution, estimating a complex mask on degraded spectrum. 
This generative branch operates at 16 kHz, since most of the speech information is concentrated in low frequencies.

\subsection{Fusion Network}
The final output is synthesized by a lightweight USES network \cite{uses} that adaptively integrates both branches. After resampling the output $\hat{\mathbf{Y}}_{gen}$ from semantic-conditioned refinement branch to match the sampling rate of 
discriminative output $\hat{\mathbf{Y}}_{disc}$,
the network estimates a time-frequency fusion mask $\mathbf{M}_{fuse}$ to compute the final spectrum as:
\begin{equation}
    \hat{\mathbf{Y}}_{final} = \mathbf{M}_{fuse} \odot \hat{\mathbf{Y}}_{disc} + (1 - \mathbf{M}_{fuse}) \odot \hat{\mathbf{Y}}_{gen},
\end{equation}
where $\odot$ denotes element-wise multiplication.

\subsection{Loss Functions}
\label{sec:loss}

For the discriminative branch, we employ the multi-resolution STFT Loss \cite{mstft}.
The generative branch is optimized using the Negative Log-Likelihood loss ($\mathcal{L}_{NLL}$) for token prediction and the Regression loss ($\mathcal{L}_{Reg}$) for mask estimation. 
$\mathcal{L}_{Reg}$ is defined as a weighted sum of the mean squared errors on the complex and magnitude spectrum, together with a perceptual loss \cite{pmsqe}:
\begin{equation}
    \mathcal{L}_{Reg} = 0.1 \cdot \mathcal{L}_{complex} + 0.9 \cdot \mathcal{L}_{mag} + 0.01 \cdot \mathcal{L}_{PMSQE}.
\end{equation}
Therefore, the loss of generative branch is formulated as 
\begin{equation}
    \mathcal{L}_{gen} = \mathcal{L}_{NLL} + \mathcal{L}_{Reg}.
\end{equation}
The fusion network is trained using  multi-resolution STFT \cite{mstft} and L1 losses with a perceptual loss: 
\begin{equation}
    \mathcal{L}_{fusion} =  \mathcal{L}_{MSTFT} + 0.5 \mathcal{L}_{L1} + \mathcal{L}_{SQA}
\end{equation}
Here, $\mathcal{L}_{\text{SQA}}$ is a score-based loss derived from the multi-metric quality assessment model~\cite{sqa}. We specifically select five perceptual metrics for supervision including MOS, DNSMOS~\cite{dnsmos}, ScoreQ~\cite{scoreq}, UTMOS~\cite{utmos}, and NISQA~\cite{nisqa}.

\section{Experiments}
\label{sec:exp}

\subsection{Experimental Setup}
We train our model on the data from Track 1 of the URGENT 2026 Challenge, excluding the pre-trained WavLM \cite{wavlm} and X-Codec \cite{xcodec} modules.
There are approximately 1.3 million clean speech utterances across five languages, and the data simulation pipeline follows that provided by the challenge.
The TF-GridNet in discriminative branch consists of 8 blocks
with a embedding size of 64 and a LSTM hidden size of 256. 
The window and hop durations are set to 20 ms and 10 ms, respectively.
There are 12 LLaMA \cite{llama} layers with the hidden dimension set to 512. The window length and hop size in DPRNN are set to 640 and 320.

\subsection{Results and Analysis}
Table \ref{tab:results} summarizes the performance on the URGENT 2026 non-blind test set. 
Discriminative (``Dis.'') branch excels in intrusive metrics but offers lower perceptual quality, whereas the semantic-conditioned refinement (``Gen.'') branch
achieves high naturalness at the cost of reduced signal fidelity. 
Our proposed hybrid method successfully integrates these strengths. 
It maintains competitive fidelity (PESQ \cite{pesq} and ESTOI \cite{estoi}) while significantly enhancing perceptual quality (DNSMOS \cite{dnsmos} and NISQA \cite{nisqa}) compared to the baseline. 
Notably, our approach also achieves the best performance in 
speaker similarity (SpkSim) and downstream speech recognition task (CAcc),
proving the efficacy of the hybrid architecture for USE.

\begin{table}[h]
\centering
\caption{Results on the URGENT 2026 Non-Blind Test set.}
\label{tab:results}
\resizebox{1.0\columnwidth}{!}{
\begin{tabular}{lcccccc}
\toprule
\textbf{Method} & \textbf{PESQ}~\cite{pesq} $\uparrow$ & \textbf{ESTOI}~\cite{estoi} $\uparrow$ & \textbf{DNSMOS}~\cite{dnsmos} $\uparrow$ & \textbf{NISQA}~\cite{nisqa} $\uparrow$ & \textbf{SpkSim} $\uparrow$ & \textbf{CAcc (\%)} $\uparrow$ \\
\midrule
Noisy            & 1.30 & 0.62 & 1.67 & 1.44  & 0.54 & 79.28 \\
Baseline         & 2.47 & 0.82 & 2.90 & 2.98 & 0.75 & 85.79 \\
Dis.       & \textbf{2.65} & \textbf{0.85} & 2.92 & 2.97 & 0.76 & 86.37 \\
Gen. & 1.96 & 0.74 & 2.98 & \textbf{3.43} & 0.73 & 84.65 \\
Hybrid  & 2.58 & 0.83 & \textbf{3.01} & 3.26 & \textbf{0.77} & \textbf{86.83} \\
\bottomrule
\end{tabular}
}
\vspace{-1em}
\end{table}

\section{Conclusion}
\label{sec:con}

In this work, 
we introduced a system submitted to the Track 1 in URGENT 2026 challenge, 
which combines discriminative precision with generative richness and naturalness. 
Experimental results on the test set demonstrate that our system is more competitive than the baseline. 
However, the generative branch is limited to 16 kHz speech and suffers from high inference latency.
Future work will focus on full-band processing and efficiency optimization for real-time deployment.

\bibliographystyle{IEEEbib}
\bibliography{refs}

\end{document}